\documentclass[onecolumn, times, trackchanges]{aastex63}

\usepackage{txfonts}

\newcommand{\target}{KIC 1573174 }

\begin{document}

\title{Frequency analysis of \target: shedding light on the nature of HADS stars}
\shorttitle{$Kepler$ Observations of \target}
\shortauthors{Lv, C. et al.}

\author{Chenglong Lv}
\affil{Xinjiang Astronomical Observatory, Chinese Academy of Sciences, Urumqi, Xinjiang 830011, People's Republic of China}
\affil{School of Astronomy and Space Science, University of Chinese Academy of Sciences, Beijing 100049, People's Republic of China}

\author{Ali Esamdin}
\email{aliyi@xao.ac.cn}
\affil{Xinjiang Astronomical Observatory, Chinese Academy of Sciences, Urumqi, Xinjiang 830011, People's Republic of China}
\affil{School of Astronomy and Space Science, University of Chinese Academy of Sciences, Beijing 100049, People's Republic of China}

\author{J.Pascual-Granado}
\affil{Instituto de Astrof\'isica de Andaluc\'ia - CSIC, 18008 Granada, Spain}

\author{Taozhi Yang}
\affil{School of Physics, Xi'an Jiaotong University, Xi'an 710049, People's Republic of China}

\author{Dongxiang Shen}
\affil{School of Physical Science and Technology, Xinjiang University, Urumqi 830046, China}


\begin{abstract}

We propose that \target is a quadruple-mode $\delta$ Scuti star with pulsation amplitudes between those of the HADS (high-amplitude Delta Scuti star) group and average low-amplitude pulsators. The radial modes detected in this star provide a unique opportunity to exploit asteroseismic techniques up to their limits. Detailed frequency analysis is given for the light curve from the Kepler mission. The variation of the light curve is dominated by the strongest mode with a frequency of F0 = 7.3975 $\rm{d^{-1}}$, as shown by Fourier analysis of long cadence data (Q1-Q17, spanning 1460 days), indicating that the target is a $\delta$ Scuti star. The other three independent modes with F1 = 9.4397 d$^{-1}$, F2 = 12.1225 d$^{-1}$ and F3 = 14.3577 d$^{-1}$, have ratios of $P_{1}$ / $P_{0}$, $P_{2}$ / $P_{0}$ and $P_{3}$ / $P_{0}$ estimated as 0.783, 0.610 and 0.515, which indicate that \target is a quadruple-mode $\delta$ Scuti star. A different approach has been used to determine the $O-C$ through the study of phase modulation. The change of period $(1/P)~dP/dt$ is obtained resulting in $-1.14 \times 10^{-6}~\text{yr}^{-1}$ and $-4.48 \times 10^{-6}~\text{yr}^{-1}$ for F0 and F1 respectively. Based on frequency parameters (i.e., F0, F1, F2, and F3), a series of theoretical models were conducted by employing the stellar evolution code MESA. The ratio of observed $f_{1}/f_{2}$ is larger than that of the model, which may be caused by the rotation of the star. We suggest high-resolution spectral observation is highly desired in the future to further constrain models.

\end{abstract}


\section{Introduction}

The high-precision photometric data provided by $Kepler$ (e.g. \citealp{2010Sci...327..977B,Koch}) provides an unprecedented opportunity to explore stellar interiors by using the natural oscillation mode of stars, thus greatly expanding the research field of asteroseismology (e.g. \citealp{Chaplin2010,Aer10,2018ApJ...861...96X,2021MNRAS.504.4039B}). The ultra-high precision photometric observations at the $\mu$ mag level have significantly advanced our understanding of several types of pulsating variable stars (e.g. \citealp{Balona2012a,2015MNRAS.452.2127S,2022MNRAS.511.1529S,2022MNRAS.tmp..536S}). \citet{Bedding2011} proposed that the observed period spacings of gravity modes could be applied to distinguish the hydrogen and helium burning stars in red giants. \citet{2018Natur.554...73G} suggested that an oxygen-dominated core may be present in pulsating white dwarfs. As a group of traditional variable stars, $\delta$ Scuti stars are excellent targets for asteroseismology research owing to their rich pulsation patterns (e.g. \citealp{2011MNRAS.414.1721B,2011MNRAS.417..591B}). The fundamental, first, second, and even third and fourth radial pulsation modes could be indicators of the internal burning mechanism of $\delta$ Scuti stars \citep{Breger2000}.

The parameter space covered by $\delta$ Scuti stars in the Hertzsprung$-$Russell (HR) diagram is of great importance for testing stellar evolution models. They cover the transition region from slowly-rotating low-mass stars with radiative cores and thick convective envelopes ($M \leq 1.5$~M$_{\odot}$) to rapidly-rotating intermediate mass stars with convective cores and predominantly radiative envelopes ($M \geq 2.5$~M$_{\odot}$). This transition in stellar structure allows many different aspects of physics to be investigated, including pulsation, rotation, magnetic fields and chemical peculiarities (e.g. \citealp{2015MNRAS.447.3948M,2015MNRAS.447.3264S,2019ApJ...887..253C,2021MNRAS.500.1992T,2021MNRAS.504.4039B}). The $\delta$ Scuti-type pulsating stars are located at the intersection of the classical Cepheid instability strip and the main sequence on the HR~diagram. $\delta$ Scuti stars typically range from A2 to F2 in spectral type with luminosity classes from III to V (e.g. \citealp{Breger2000,1990A&AS...83...51L,2001A&A...366..178R}), and within the effective temperature range of 6300K $\leq$ T$_{\rm eff}$ $\leq$ 8600K \citep{2011A&A...534A.125U}. They pulsate mainly in radial and non-radial modes (e.g. \citealp{Breger2000,2011A&A...534A.125U}) and are typically excited in the $\kappa$ mechanism (e.g. \citealp{Breger2000,Aer10}), these pulsation modes are generally identified as low radial-order ($n$) low-degree ($l$) pressure ($p$) modes (e.g. \citealp{1998A&A...335..549V,Aer10,2011A&A...534A.125U,2019ApJ...887..253C}). They are also found in binary systems (e.g. \citealp{2019ApJ...885...46G,2020MNRAS.498.4272M,2021MNRAS.505.3206M,2021RAA....21..224L}) and thus, these targets are excellent samples for asteroseismic study, as they could improve our understanding of stellar structure and evolution.

HADS (high-amplitude Delta Scuti star) stars are a subclass of $\delta$ Scuti stars with peak-to-peak light amplitudes larger than 0.3 mag. They are traditionally found to be slow rotators with $v$ sin $i$ $<$ 30 km s$^{-1}$ and pulsation periods between 1 and 6 h \citep{McNamara2000}. From the ground-based observations, HADS typically have only one or two radial pulsation modes in the fundamental and/or first overtone mode (e.g. \citealp{Yang2012,Niu2017,2018ApJ...861...96X,Yang2018}). In recent decades, thanks to the high photometric precision observations from space telescopes, especially in $Kepler$ mission \citep{2010Sci...327..977B}, and the development of sophisticated data analysis techniques such as \citet{Lares}, low amplitude frequencies can also be found in the frequency spectrum of HADS \citep{2021MNRAS.504.4039B}. By using $Kepler$ data, three independent frequencies were identified in KIC 10975348 as radial modes, which reclassified this star as a triple-mode HADS \citep{2021AJ....161...27Y}. \citet{2021MNRAS.504.4039B} analyzed the light variation of KIC 5950759, and 12 additional independent frequencies were extracted but regarded as non-radial modes. \citet{2021AJ....162...48L} report a detailed light-curve analysis of the Kepler target KIC 12602250, and their results show that KIC 12602250 is just pulsating at two radial frequencies. Therefore, detecting low-amplitude frequencies will enrich the features of light variation and improve the understanding of HADS.

HADS stars typically have only one or two radial pulsation modes of the fundamental and/or first overtone mode. Recently, there have been several studies of triple-mode stars but no quadruple-mode star has been found to this date. \citet{2021arXiv211013594Y} list about 155 radial double-mode HADS. Radial triple-mode HADS are particularly rare, \citet{2008wils} list only four known HADS that pulsate in three radial modes simultaneously. Additional four radial triple-mode HADS have been discovered in recent years thanks to high-precision data from space telescopes (e.g. \citealp{2016AJ....152...17M,2021AJ....161...27Y,2021ApJ...922..199S,2021A&A...655A..63Y}). Comparisons of single-mode, double-mode, triple-mode and quadruple-mode HADS may illuminate what determines the number of radial modes a pulsating star has, something that is still not well-understood. The discovery of additional multi-mode radial pulsators would greatly help with these comparisons. More radial pulsation modes can better constrain the results of the fitting models, so using the four radial modes detected in the HADS star we analyze in this work we have a unique opportunity to exploit the asteroseismic techniques to their limits.

\target ($\alpha_{2000}$=19$^{h}$:25$^{m}$:28.8$^{s}$, $\delta_{2000}$=+37$\degr$:09$\arcmin $:23.7$\arcsec$) is classified as a $\delta$ Scuti star with a pulsation period of 3.24 hrs by \cite{{2011A&A...529A..89D}}. In this work we propose for that \target is a relatively large-amplitude radial pulsator radial quadruple-mode $\delta$ Scuti star by frequency analysis of its amplitude spectrum. The fundamental parameters of this star are listed in Table \ref{Tab1}.

In this paper, Section 2 introduce the observations of \target. The frequency analysis is presented in Section 3, and the $O-C$ analysis of the star is presented in Section 4. In Section 5, we construct stellar evolution models and make pulsation frequency fitting. A brief discussion and the conclusions are presented in Sections 6 and 7, respectively.

\begin{deluxetable}{ccc}
\renewcommand\arraystretch{1.2}
\tabletypesize{\small}
\setlength\tabcolsep{20pt}
\tablewidth{0.5\textwidth}
\tablenum{1}
\tablecaption{\target observational (photometry) data characteristics \label{Tab1}}
\tablehead{
\colhead{Parameters} &
\colhead{Value in \href{https://kasoc.phys.au.dk/catalog/12602250}{Catalog}}  &
\colhead{ }
}
\startdata
Kmag                  & 13.551                                & a           \\
RA       & 19$^{h}$:25$^{m}$:28.8$^{s}$                       & a           \\
Dec      &+37$\degr$:09$\arcmin $:23.7$\arcsec$               & a           \\
BJD$_\mathrm{0}$      & 2454964.5126                          & a           \\
Period                & 3.24     hr                           & a           \\
$T_\mathrm{eff}$      & 6971     K                            & a           \\
                      & 7390     $\pm$ 150 K                  & b           \\
log $g$               & 4.035    dex                          & a           \\
                      & 3.929    dex                          & b           \\
$\frac{R}{R_{\sun}}$  & 1.813                                 & a           \\
$\frac{Fe}{H}$        & -0.149                                & a           \\
                      & 0.1825                                & b           \\
Parallax (mas)        & 0.372  $\pm$ 0.016                    & c           \\
Rayleigh$~f_\mathrm{res}$& 0.001  d$^{-1}$                                  \\
   B     &  14.229      &   d \\
   V     &  13.572      &   d \\
   J     &  12.616      &   d \\
   H     &  12.472      &   d \\
   K     &  12.417      &   d \\
   \enddata
   \tablecomments{(a) Parameters from the KASOC. (b) LAMOST http://dr7.lamost.org/. (c) Gaia \citep{2017MNRAS.471..770M}. (d) TASOC https://tasoc.dk/catalog/.}
\end{deluxetable}

\begin{figure*}
\begin{center}
  \includegraphics[width=0.85\textwidth]{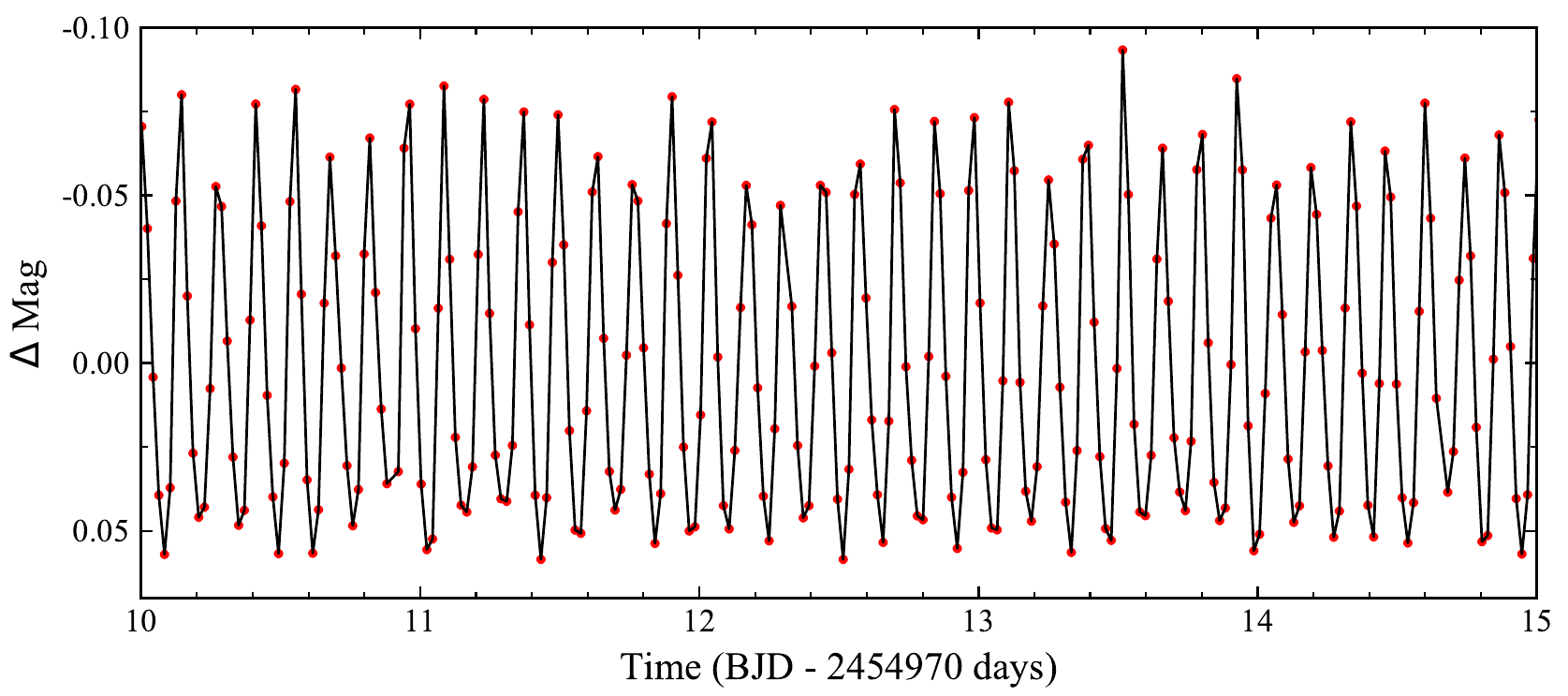}
  \caption{A portion of the long cadence light curve of \target. The amplitude of the light curve is about 0.15 mag.}
    \label{fig:light curve}
\end{center}
\end{figure*}

\section{OBSERVATIONS AND DATA REDUCTION}

The $Kepler$ Space Telescope observed \target for seventeen quarters (i.e., Q1-Q17) from BJD 2454964.513 to 2456424.001. Through the Kepler Asteroseismic Science Operations Center (KASOC) database\footnote{KASOC: {https://kasoc.phys.au.dk/search/}}, only long cadence (LC) photometric observations are available for \target. The two types of data are the raw flux, reduced by the NASA Kepler Science Pipeline, and the corrected flux, provided by the KASOC Working Group 4 (WG \# 4: $\delta$ Scuti targets), respectively \citep{Kjeldsen2010}. We use the corrected flux and convert it to magnitude. The average value for each quarter is then subtracted to obtain the corrected time series, as the second type of data has been corrected for systematic errors such as the cooling down, warming up, outliers, and jumps. After the above processing, a rectified light curve of 64,795 data points with a time span of about 1460 days was finally obtained.

Some stars with peak-to-peak amplitudes lower than 0.3 mag behave like HADS. \citet{2011MNRAS.414.1721B} studied KIC 9700322 using $Kepler$ data and found that the star has a slow rotation and dominant radial modes, typical of HADS pulsations, but with a peak-to-peak amplitude lower than the 0.3 mag that initially defined of the HADS. Figure \ref{fig:light curve} shows a portion of the rectified light curve of \target covering five days. From this figure, the peak-to-peak amplitude of \target obtained from the rectified light curve is $\sim$0.15 mag, so this star is in the transition region between small amplitude and high amplitude $\delta$ Scuti stars.

\section{FREQUENCY ANALYSIS}

In order to analyze the frequencies present in the light curve, we use the software PERIOD 04 \citep{Lenz2005} to analyze the pulsating behavior of \target.

\begin{figure*}
\begin{center}
  \includegraphics[width=0.85\textwidth]{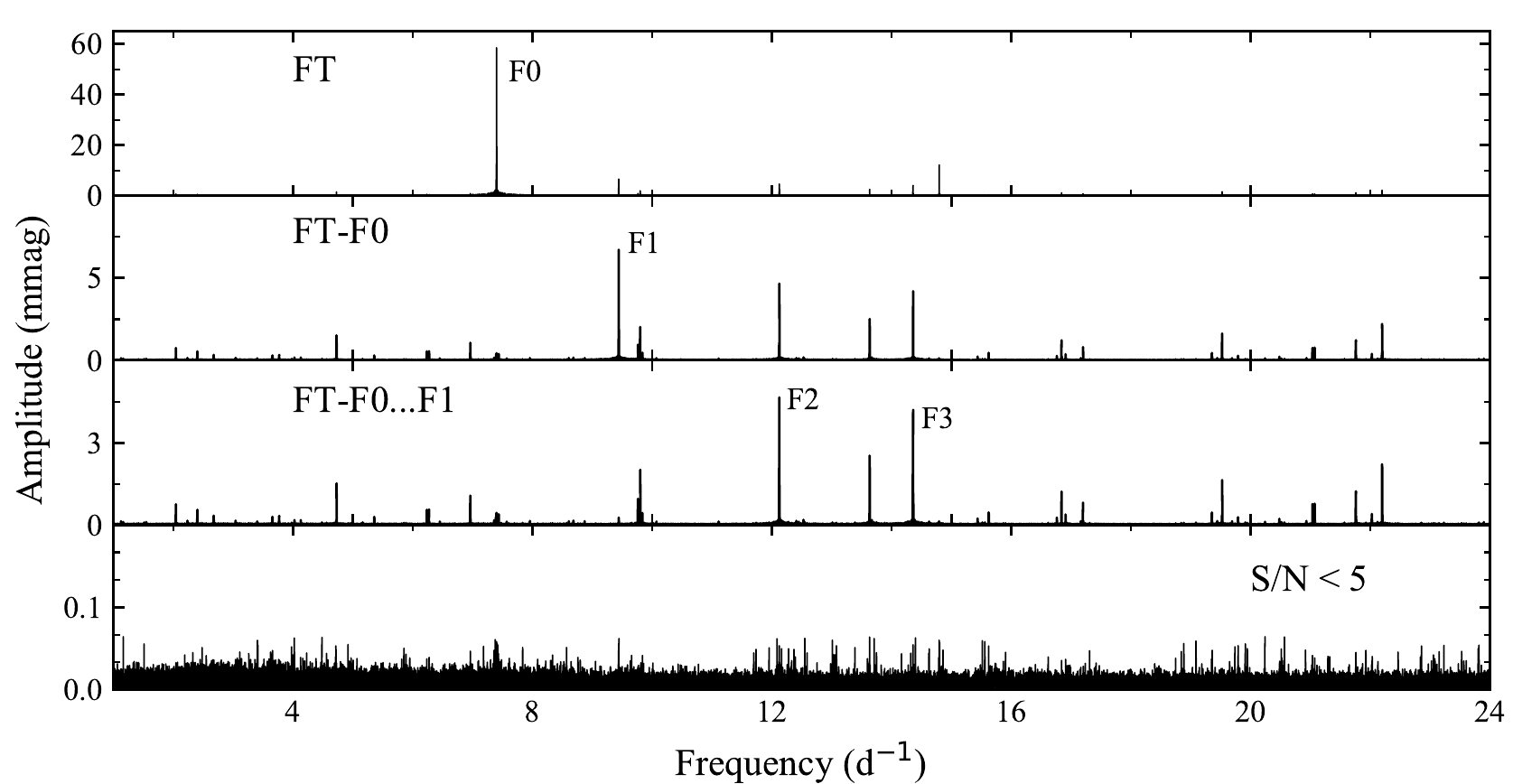}
  \caption{Fourier amplitude spectra and the prewhitening process for the light curve of \target. The top panel shows the fundamental frequency F0. The two middle panel show the other three independent frequencies, F1, F2, and F3, respectively. The bottom panel shows the residual after subtracting 54 significant frequencies and the S/N $<$ 5.}
    \label{fig:amplitude spectra}
\end{center}
\end{figure*}

The Nyquist frequency of LC observations is $f_{N}$ = 24.469 d$^{-1}$, so the frequency is limited to the range of 0 $<$ $f$ $<$ 24 d$^{-1}$ during our analysis. We use the resolution frequency $f_{res}$ = 1 / $T$ to distinguish two frequencies that are very close to each other, and if the difference between these two frequencies is greater than the resolution frequency, we consider these two frequencies to be resolved. The resolution frequency $f_{res}$ = 1 / $T$ is 0.00068~d$^{-1}$ for \target LC light curve. In the process of extracting significant frequencies, we usually identify the highest peaks as significant frequencies. The rectified light curve was fitted with the following formula:
\begin{equation}
m = m_{0} + \sum_{\mbox{\scriptsize\ $i$=1}}^N\mathnormal{A}_{i}sin(2\pi(\mathnormal{f}_{i}\mathnormal{t} + \phi_{i})), \label{equation1}
\end{equation}
where $m_{0}$ is the zero-point, $A_{i}$ is the amplitude, $f_{i}$ is the frequency, and $\phi_{i}$ is the corresponding phase. The multi-frequency least square fit of the light curve for all detected significant frequencies is then performed using Eq.\ref{equation1} to obtain solutions for all frequencies. The residuals are obtained by subtracting the theoretical light curve constructed by the above solution from the rectification data and continuing the next search using the obtained residuals, repeating the above steps until no significant peaks are found in the spectrum. As the criterion for determining the significance of the detected peaks, we utilize the S/N $>$ 5 suggested by \cite{2015MNRAS.448L..16B}. The adoption of this higher detection threshold does not affect the results of the frequency analysis for this star. The frequency uncertainty was determined according to the method proposed by \citet{1999DSSN...13...28M}. Figure \ref{fig:amplitude spectra} shows the amplitude spectra and the prewhitening procedures of the light curve. The top panel shows the fundamental frequency F0, and the two middle panels show the other three independent frequencies, F1, F2, and F3, respectively. The residuals after subtracting the 54 significant frequencies are shown in the bottom panel. No significant peaks with S/N $>$ 5 in the residual spectra can be detected, showing the overall distribution of typical noise.

\begin{deluxetable}{cccccc}
\renewcommand\arraystretch{1.2}
\tabletypesize{\footnotesize}
\setlength\tabcolsep{2.5pt}
\tablewidth{0.5\textwidth}
\tablenum{2}
\tablecaption{The radial pulsation mode frequencies in LC data of \target and their significant harmonics and combination frequencies. \label{Tab2}}
\tablehead{
\colhead{$f_{i}$}   &
\colhead{Frequency (d$^{-1}$)}  &
\colhead{Amplitude (mmag)}      &
\colhead{S/N}            &
\colhead{Comment} &
}
\startdata
1&	7.3975216(3)&	58.58(4)&	3379.3	&	F0	\\
2&	9.439709(3) &	 6.74(4)&	461.3	&	F1	\\
3&	12.122497(4)&	 4.65(4)&	334.8	&	F2	\\
4&	14.357835(2)&	 4.19(4)&	635.8	&	F3	\\
5&	14.795043(1)&	12.21(4)&	863.7	&	2F0	\\
6&	22.192564(8)&	 2.21(4)&	165.1	&	3F0	\\
7&	19.52002(1) &	 1.63(4)&	121.9	&	F0+F2	\\
8&	4.72496(1)  &	 1.51(4)&	88.0    &	F2-F0	\\
9&	21.75536(1) &	 1.31(4)&	98.4    &	F0+F3	\\
11&	16.83723(2) &	 1.21(4)&	89.6    &	F0+F1	\\
12&	6.96031(2)  &	 1.12(4)&	64.4    &	F3-F0	\\
13&	2.04215(4)  &	 0.75(4)&	42.2    &	F1-F0	\\
14&	24.2347(1)  &	 0.42(4)&	30.6    &	2F0+F1  \\
15&	9.83632(5)  &	 0.42(4)&	29.2    &	2F3-2F1	\\
16&	2.67257(5)  &	 0.34(4)&	18.1    &	2F0-F2	\\
17&	5.35536(6)  &	 0.30(4)&	18.1    &	2F0-F1	\\
18&	0.25426(6)  &	 0.32(4)&	11.8    &	3F2-F0-2F3	\\
19&	4.02277(9)  &	 0.18(4)&	9.9     &	2F3-F1	\\
20&	8.6061(1)   &	 0.13(4)&	9.0     &	4F1-2F0-F3	\\
21&	10.0700(1)  &	 0.13(4)&	9.0     &	3F0-F2	\\
22&	2.2353(1)   &	 0.16(4)&	8.8     &	F3-F2	\\
23&	2.6828(1)   &	 0.10(4)&	5.5     &	F2-F1	\\
   \enddata
    \tablecomments{Among these frequencies, 4 peaks are independent frequencies, others are harmonic or combinations (denoted by $f_{i}$).}
\end{deluxetable}

\begin{figure*}
\begin{center}
  \includegraphics[width=0.45\textwidth]{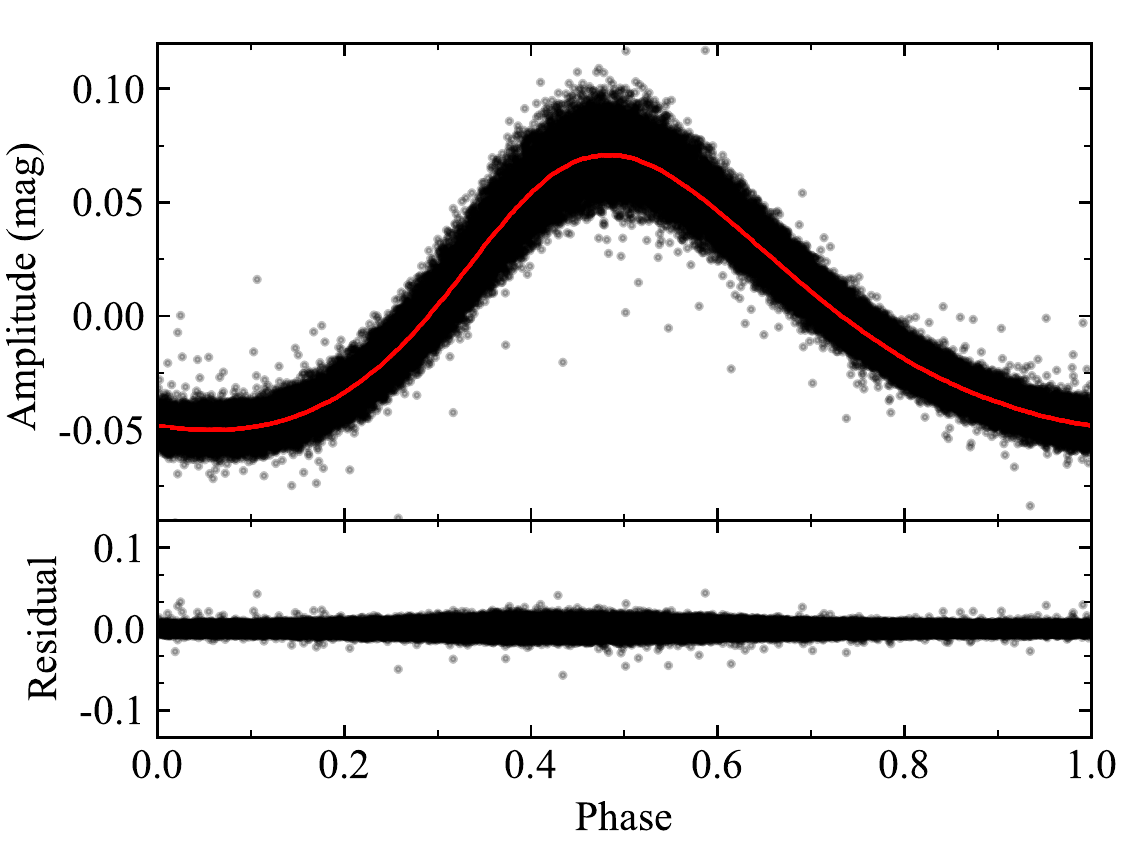}
  \caption{Phase diagram of \target, folded by the fundamental frequency F0 = 7.3975216(3) d$^{-1}$. The phase diagram shows that the light variation has a tendency to climb rapidly and fall slowly, which is typical of HADS.}
    \label{fig:phase curve}
\end{center}
\end{figure*}

A total of 54 frequencies detected by performing Fourier transformation from the spectrum of \target. Combination frequencies were identified by searching for linear sum and difference frequencies, n$\nu$$_{i}$ $\pm$ m$\nu$$_{j}$, with the \citet{1978Ap&SS..56..285L} criterion as a tolerance and assuming that the highest-amplitude peaks within a combination family are the real pulsation mode frequencies \citep{2015MNRAS.450.3015K}. Among them, four radial modes and the combination and harmonics frequencies of the four radial modes are listed in Table \ref{Tab2}. and the remaining 31 are non-radial pulsation frequencies listed in Table \ref{TabA}. From these frequencies, four are considered independent. The two high-amplitude modes have a period ratio of 0.783 identifying them as the fundamental and first overtone radial modes \citep{2000ASPC..210.....B}. Therefore, we marked $f_{1}$ with 'F0' in the last column of Table \ref{Tab2}. \citet{{Stellingwerf1979}} was one of the first papers to predict the period ratios based in theoretical structure models and presented the period ratios of the first four radial modes as: $P_{1}$ / $P_{0}$ = (0.756 - 0.787), $P_{2}$ / $P_{0}$ = (0.611 - 0.632) and $P_{3}$ / $P_{0}$ = (0.500 - 0.525), in which $P_{0}$, $P_{1}$, $P_{2}$ and $P_{3}$ represent the fundamental mode, first overtone, second overtone and third overtone, respectively. The ratio of $P_{1}$ / $P_{0}$, $P_{2}$ / $P_{0}$ and $P_{3}$ / $P_{0}$ of \target are measured as 0.783, 0.610 and 0.515, respectively. So the other three independent frequencies $f_{2}$, $f_{3}$ and $f_{4}$ are labeled as 'F1', 'F2', and 'F3', respectively. In addition, some harmonic (i.e., $f_{5}$, $f_{6}$) of 'F0' and a lot of combination frequencies of 'F0', 'F1', 'F2', 'F3' are also detected. We note that although these frequencies have been detected previously by \citet{2016MNRAS.460.1970B}, it is the first time that they have been identified as radial modes in our study. Figure \ref{fig:phase curve} shows the phase diagram of \target, folded by the fundamental frequency F0 = 7.3975216(3)~d$^{-1}$. The phase diagram shows that the light variation has a tendency to climb rapidly and fall slowly, which is typical of HADS.

\section{$O - C$ Diagram}
An $O-C$ diagram can be used to calculate period changes in pulsation modes \citep{Percy80, Breger98}. It consists in studying the observed deviation of the times of maximum (or minimum) light from the calculated times assuming a constant period. In this section we outline this technique and describe its application to the light curve of \target.

In \citet{Percy80} it is shown that when the true period is changing at a uniform rate, the phase-shift diagram is an upward (downward) parabola for increasing (decreasing) period. Here phase $\phi_i$ means the decimal part of the cycle number $f_i = \Delta T_i/P$ with P the period, $\Delta T_i = (T_i-T_0)$ the time lapse between $T_i$, the observed time of a maximum, and $T_0$ the reference time (e.g. the first maximum). The equation for the cycle number $f_i$ is then, according to \citet{Percy80}, given by:
\begin{equation}
\label{phaseeq}
 f_i = \left( \frac{P_0}{P} \right) E_i +  \frac{1}{2} \left(\frac{dP}{dt}\right) E_i^2
\end{equation}
where $E_i$, the independent variable, is just the integer number of cycles counted from $T_0$ to $T_i$ and $P_0$ is the assumed period (e.g. mean period).

Substituting the expression for $f_i$ in the previous equation and solving for $T_i$:
\begin{equation}
\label{omc}
\large T_i = T_0 + P_0 E_i +  \frac{1}{2} \beta E_i^2
\end{equation}
where $\beta=P\left(\frac{dP}{dt}\right)$.

In Eq.\ref{omc} the first two terms that are summed on the right hand side are just the maxima calculated for constant period $P_0$, that is, the calculated times of maximum or minimum light are $T_0 + P_0 E_i$  while the observed are $T_i$. Finally the difference $O-C$ is:
\begin{equation}
\label{omc1}
O-C = \frac{1}{2}~\beta E_i^2 = 0.5\left(\frac{1}{P}\frac{dP}{dt}\right) \Delta T_i^2
\end{equation}

Either Eq.\ref{omc} or \ref{omc1} can be used to obtain the period change by fitting a parabola for a number of observed maxima. We can obtain a precise determination of $(1/P)dP/dt$, which is the rate of change of the period. This is converted to the standard units of $yr^{-1}$ by multiplying by a factor 365.25.

\begin{figure}
\begin{center}
  \includegraphics[width=0.45\textwidth]{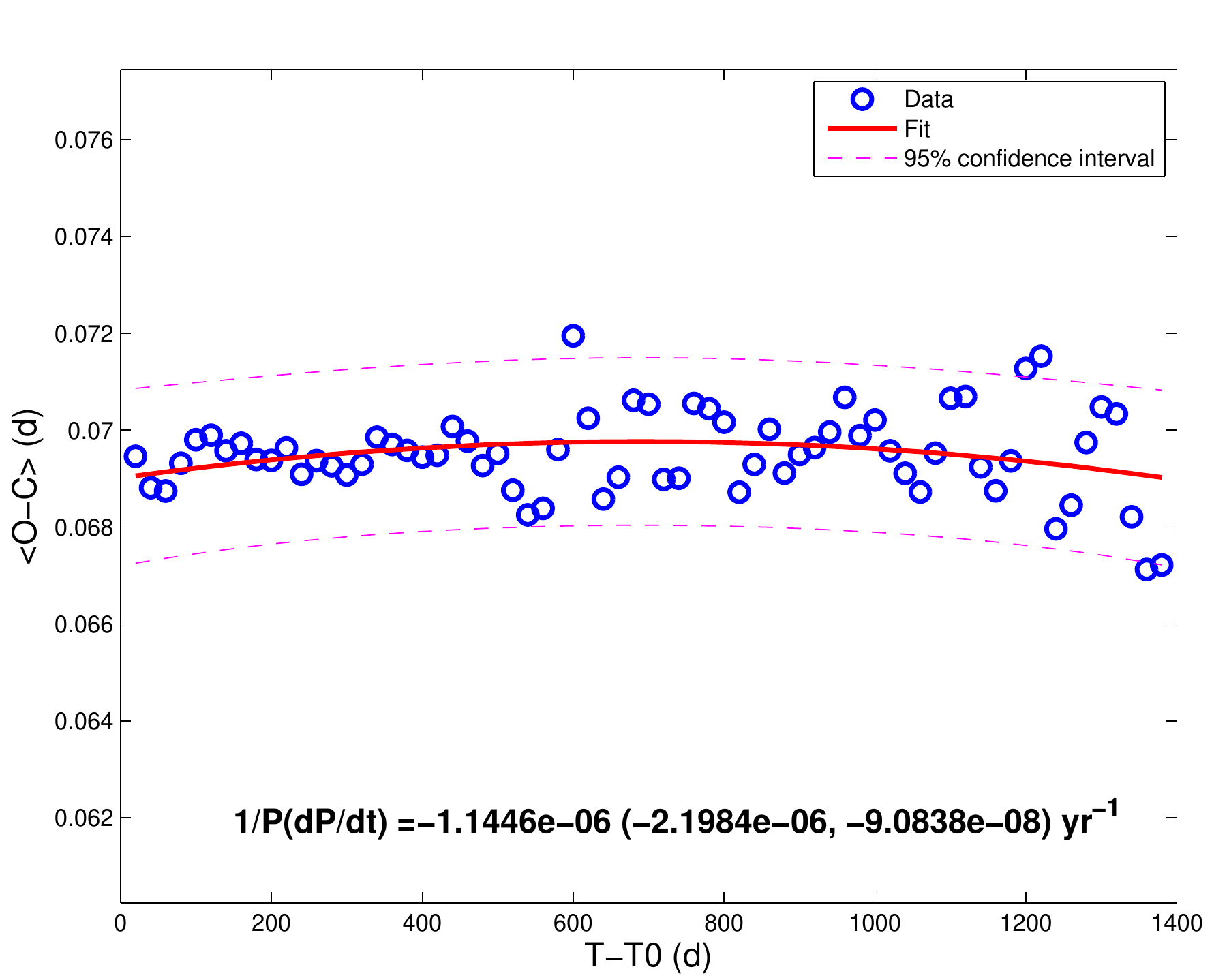}
  \includegraphics[width=0.45\textwidth]{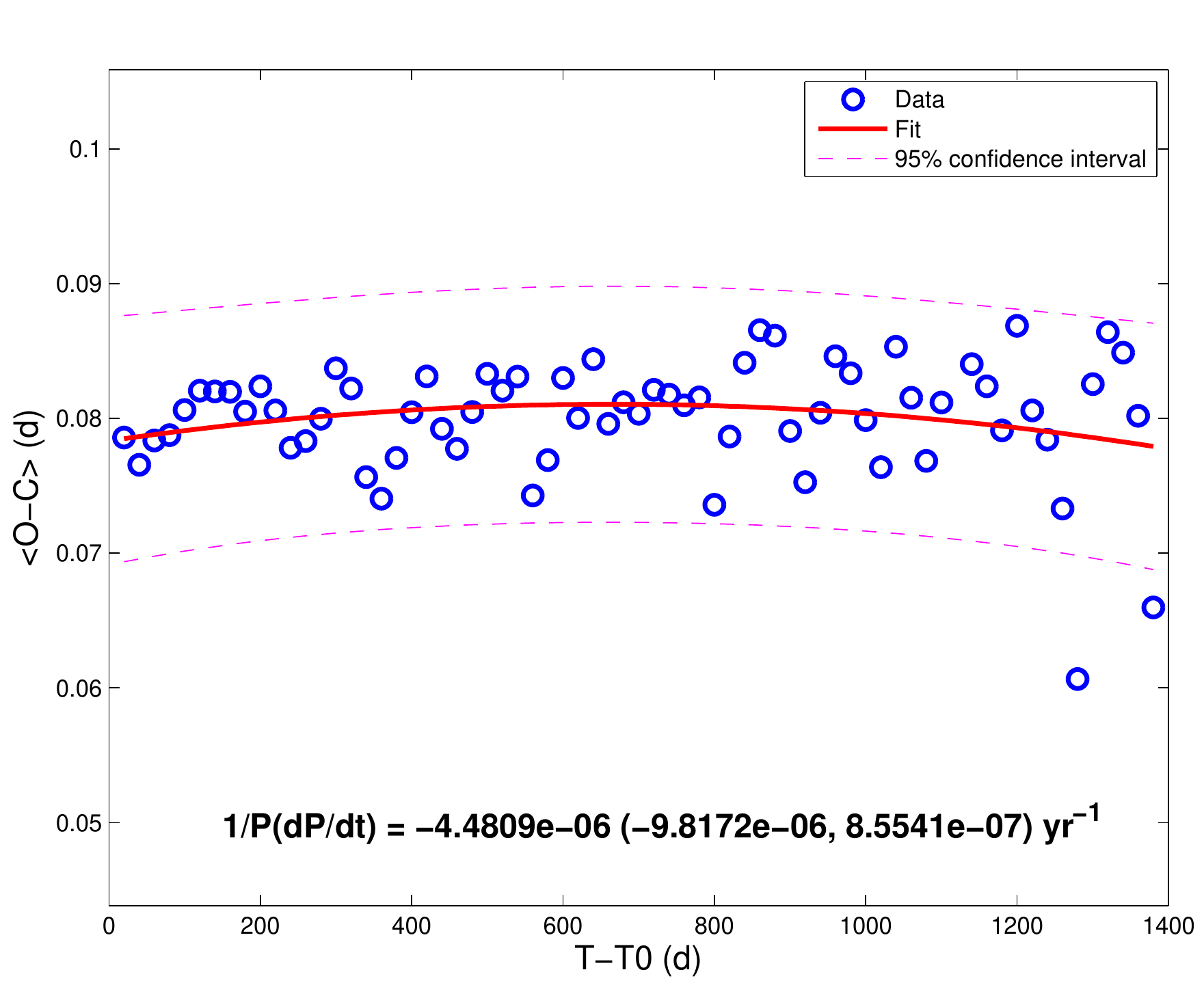}
  \caption{$O-C$ diagram for the fundamental F0 (top panel) and first overtone F1 (bottom panel) of \target, based on the phases obtained from the frequency analysis of 69 bins of 100-d with 80-d overlaps. A fit of a parabola is also shown indicating a negative change of period.}
    \label{fig:omc2}
\end{center}
\end{figure}

The times of maximum light are often used in $O-C$ diagrams, but the times of maximum light showed a very high dispersion (probably caused by the presence of multiple significant frequencies), so that it is challenging to study period changes with this approach. A different approach has been used to determine the $O-C$ through the phase modulation as originally introduced by \citet{Percy80} (see Eq.\ref{phaseeq}).

The phase of a pulsation mode at a given time can be converted into the time delays $O-C$ \citep{2017ampm.book.....B} using the following expression:
\begin{equation}
    O-C = \frac{\phi(t)}{2\pi f}
\end{equation}
with the phase $\phi(t)$ in rad units. In order to study phase modulation, a multisine fitting solution provided by SigSpec algorithm is obtained for different time intervals. This algorithm is based on the dependence of the False Alarm Probability on frequencies and phases. It takes into  account the impact of the phase coverage of measurements in the distribution in Fourier space so it is more appropriate than a simple Discrete Fourier Transform (DFT) for this purpose since its frequency accuracy is much higher \citep[for more details see][]{Reegen}, so phases will be better estimated too through the multi-sinusoid fitting process. In this way phases are calculated for overlapping segments of 100 days at steps of 20 days. The resulting $O-C$ values are shown for the fundamental and first overtone radial mode in Figure \ref{fig:omc2}. Fitting a parabola to these, the rate of change of period $(1/P)~dP/dt$ is obtained resulting in $-1.14 \times 10^{-6}~\text{yr}^{-1}$ and $-4.48 \times 10^{-6}~\text{yr}^{-1}$ for F0 and F1 respectively. As illustrated in Figure \ref{fig:omc2}, the O-C values are inside the 95\% confidence intervals centered on the estimated rate of change. Also, these results appears to be consistent since the dP/dt of the first overtone is of the same order and sign as that the fundamental mode. However, since the ranges are too broad, even a complete lack of change of period might be compatible with these confidence intervals. Therefore, we conclude that it is not possible to estimate a change of period with enough statistical significance with the available data.

\section{Stellar Model and Fitting Results}

We constructed a grid of evolutionary models of stars and calculated their corresponding adiabatic frequencies using the Modules for Experiments in Stellar Astrophysics (MESA v11701; \citealp{Paxton2011, Paxton2013, Paxton2015,Paxton2018,Paxton2019}) and the stellar oscillation code GYRE \citep{2013MNRAS.435.3406T}. Our theoretical models are constructed on basis of the OPAL opacity table GS98 \citep{1998SSRv...85..161G} series. The classical mixing length theory of \citep{1958ZA.....46..108B} with $\alpha$ = 1.90 \citep{Paxton2013} is used in the convective region. Effects of element diffusion, convective overshooting, and rotation are not included in our calculations.

In our calculations, we fix the mixing-length parameter $\alpha$ = 1.90 and set the initial helium fraction $Y = 0.249 + 1.33Z$ \citep{2018MNRAS.475..981L} as a function of the metallicity $Z$. There are large uncertainties associated with determining the metallicity with a low-resolution spectrum from LAMOST \citep{2012RAA....12.1197C}. Therefore, we chose to survey a range of models with different metallicities between 0.002 $\leq$ $Z$ $\leq$ 0.030 in steps of 0.002, and determine the best-fitting mass and age. The stellar mass $M$ varies from 1.50 $M_{\odot}$ to 2.80 $M_{\odot}$ with a step of 0.01 $M_{\odot}$. Each model in the above grid was evolved from the zero-age MS to the post-MS stage with effective temperature of Teff = 5600 K.

\begin{figure}
\begin{center}
  \includegraphics[width=0.45\textwidth]{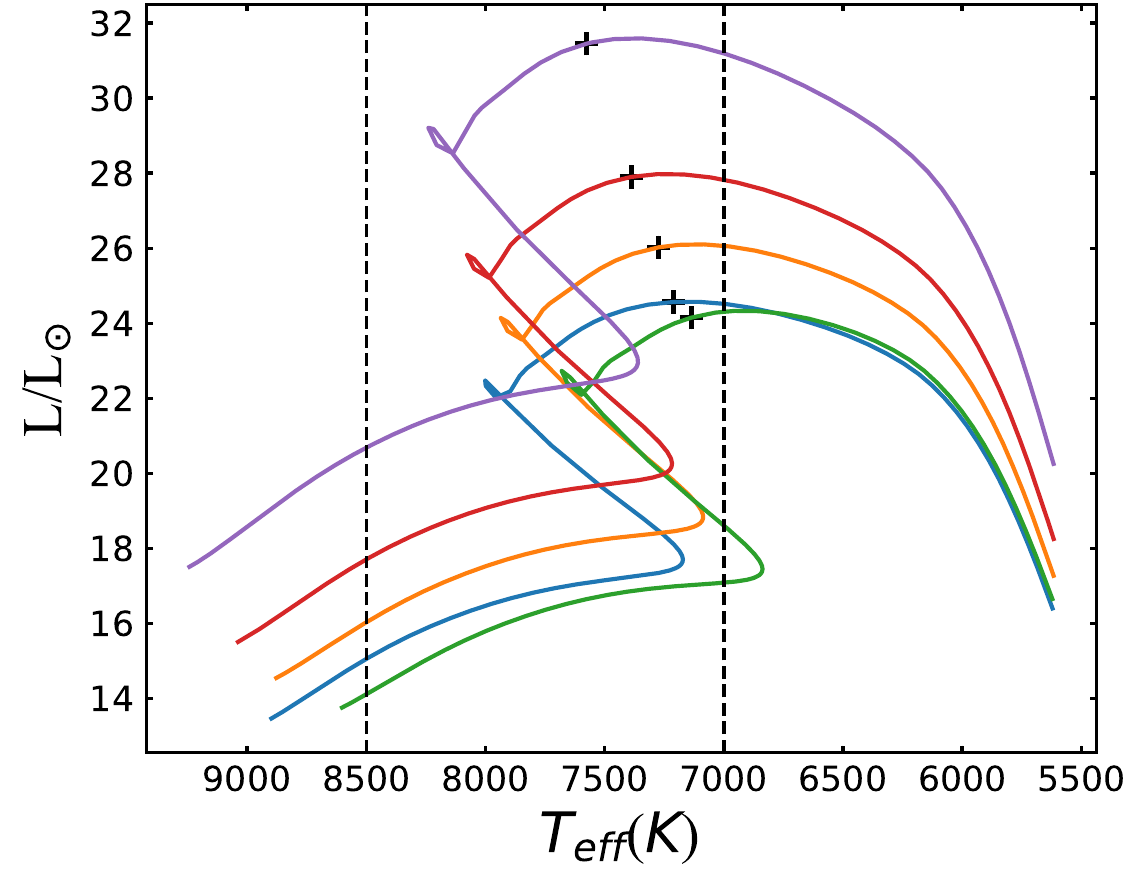}
  \caption{Evolutionary tracks from the zero-age MS to the post-MS for the 5 candidate models, as listed in Table~\ref{Tab3}. The plus sign mark the minimum $\chi^{2}$ for each specific model by fitting the calculated $f_{1}$, $f_{2}$, $f_{3}$ and $f_{4}$ with the observed values. The two vertical dotted lines mark Teff at 7000 K and 8500 K, respectively.}
    \label{fig:models}
\end{center}
\end{figure}

\begin{figure}
\begin{center}
  \includegraphics[width=0.5\textwidth]{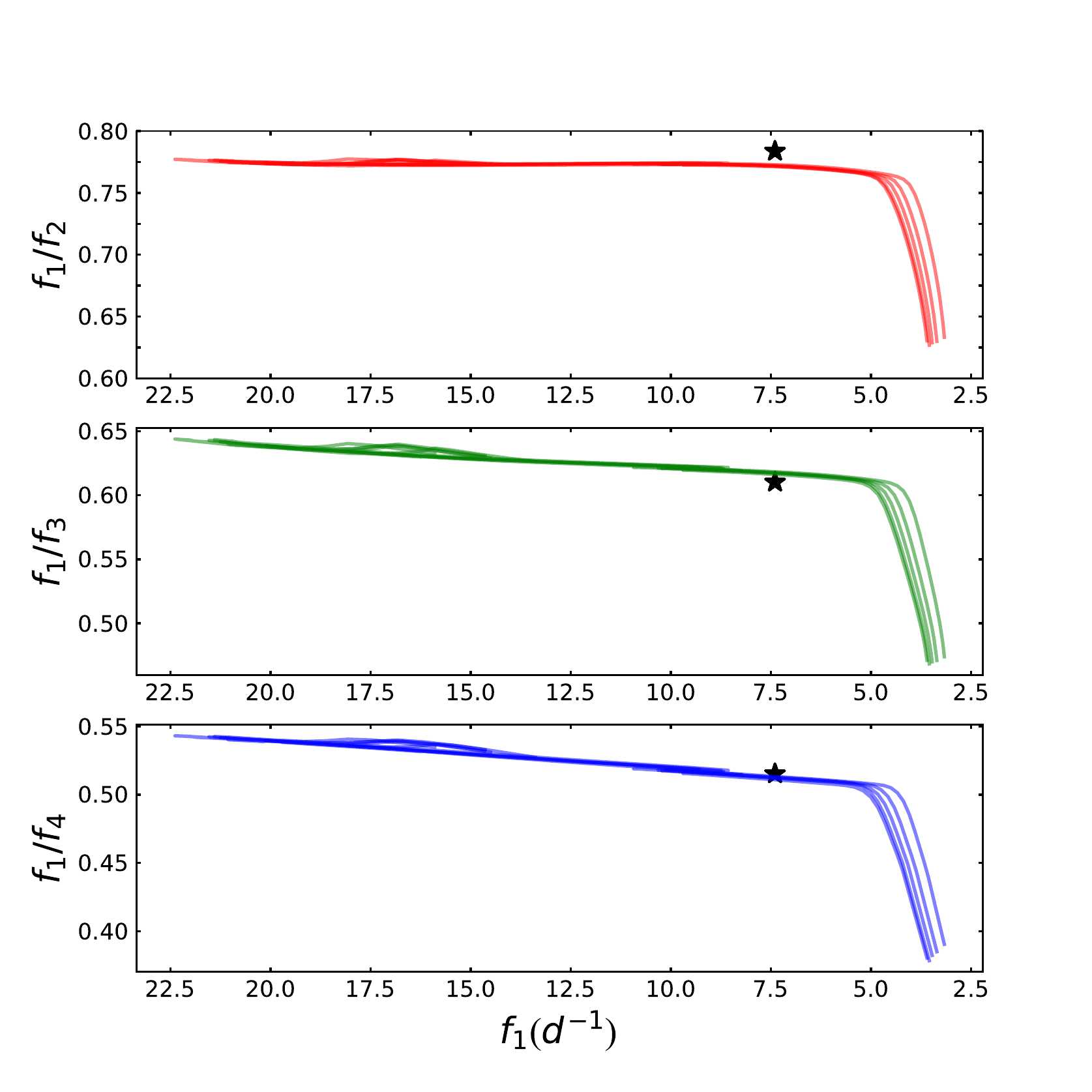}
  \caption{Petersen diagrams spanning from the zero-age MS to the post-MS phase. The black stars represent the observed ratios. From top to bottom, the ratio of the fundamental frequency to the first overtone, the ratio of the fundamental frequency to the second overtone, and the ratio of the fundamental frequency to the third overtone are shown successively.}
    \label{fig:modelsratio}
\end{center}
\end{figure}

Since $Kepler$ data are of exceptionally high quality, the four dominant pulsation mode frequencies and their frequency ratio are known to a very high precision. The high amplitudes of the pulsation modes and their frequency ratio indicate that they are likely the fundamental, first overtone, second overtone, third overtone radial modes, so a sensible method to model \target is by using Petersen diagrams. The frequency and frequency ratio of radial modes depend primarily on the mass, age, evolutionary stage, and metallicity. The Petersen diagrams therefor provides a useful diagnostic method in terms of constraining these parameters of radial pulsators (see e.g. \citealp{1973A&A....27...89P,Petersen1996,2020MNRAS.499.3034D,2021MNRAS.504.4039B}). The degeneracy between mass and age cannot generally be broken by fitting the two pulsation modes alone \citep{2021MNRAS.504.4039B}. We therefore added additional information, such as fitting the third and fourth modes. Then by using the method from (\citep{2019ApJ...887..253C}, i.e. Equation (5), $\chi^{2}$ method) to select the best-fitting models, the goodness of fit can be obtained, by comparing model frequencies with the observed frequencies F0, F1, F2, and F3. we chose a threshold of $\chi^2$ = 0.0115, since performing multiple radial frequencies fitting (four radial frequencies) and other factors, such as rotation, affects the final fitness (see Section 6), we use this $\chi^2$ value to select the best model for the region of parameter convergence. Our results are shown in Figure \ref{fig:models}, and we provide the best-fitting mass, $Z$, effective temperature, luminosity, surface gravity, age, theoretical frequency of fundamental, first overtone, second overtone, third overtone radial modes are listed in Table~\ref{Tab3}, and the frequency ratio of the fundamental and first overtone, second overtone, third overtone radial modes are shown in Figure \ref{fig:modelsratio} and listed in Table~\ref{Tab4}.

\setcounter{table}{2}
\begin{table*}[!t]
\renewcommand\arraystretch{1.2}

\caption{Candidate models with $\chi^2$ $\leq$ 0.0115.}
\label{Tab3}
\tabletypesize{\tiny}
\centering
\begin{tabular}{ccccccccccccc}
\hline
\hline
$M$ $(M_{\odot})$ &  $Z$  & $\log T_\mathrm{eff}$   & $\log(L/L_{\odot})$  &$\log g$   &Age ($10^{9}\ \mathrm{years}$)  &  $f_{1}$ (d$^{-1}$) & $f_{2}$ (d$^{-1}$) &  $f_{3}$ (d$^{-1}$) & $f_{4}$ (d$^{-1}$) &$\frac{\rm{d}P_{0}}{\rm{d}t}$ ($\times$ 10 $^{-8})$ & $\chi^{2}$  \\
\hline
1.88 & 0.016 & 3.8580 & 1.3904 & 3.7070 & 1.0918 &  7.3838   & 9.5695  &  11.9582  &14.4009  &1.24  &0.01147\\
1.95 & 0.019 & 3.8618 & 1.4154 & 3.7131 & 1.0320 &  7.3823   & 9.5660  &  11.9613  &14.4151  &1.02  &0.01136\\
1.96 & 0.024 & 3.8535 & 1.3829 & 3.7148 & 1.0709 &  7.3690   & 9.5560  &  11.9584  &14.4260  &8.10  &0.01148\\
1.97 & 0.018 & 3.8686 & 1.4456 & 3.7144 & 0.9847 &  7.3914   & 9.5723  &  11.9654  &14.4160  &0.98  &0.01142\\
2.03 & 0.018 & 3.8794 & 1.4977 & 3.7187 & 0.9025 &  7.3980   & 9.5728  &  11.9635  &14.4124  &0.90  &0.01150\\
\hline
\end{tabular}
\end{table*}

\setcounter{table}{3}
\begin{table*}[!t]
\renewcommand\arraystretch{1.2}

\caption{The frequency ratio of the theoretical fundamental and theoretical first overtone, second overtone third overtone radial modes.}
\label{Tab4}
\tabletypesize{\small}
\centering
\begin{tabular}{ccccc}
\hline
\hline
$M$ $(M_{\odot})$ &  $f_{1}/f_{2}$  & $f_{1}/f_{3}$   & $f_{1}/f_{4}$\\
\hline
1.88 & 0.7716 & 0.6175 & 0.5127 \\
1.95 & 0.7717 & 0.6172 & 0.5121\\
1.96 & 0.7711 & 0.6162 & 0.5108\\
1.97 & 0.7721 & 0.6177 & 0.5127\\
2.03 & 0.7728 & 0.6184 & 0.5133\\
\hline
\end{tabular}
\end{table*}

\section{Discussion}

The period changes due to stellar evolution for stars in and across the lower part of the classical instability strip allow an observational test of stellar evolution theory, assuming that other physical reasons for period changes can be excluded \citep{Breger2000}. From a theoretical point of view, an evolutionary change in $T_{eff}$ and $M_{bol}$ leads to a period change of size \citep[equation 9 of][]{Breger2000}
\begin{equation}
\frac{1}{P}\frac{dP}{dt}=-0.69\frac{dM_{bol}}{dt}-\frac{3}{T_{eff}}\frac{dT_{eff}}{dt}+\frac{1}{Q}\frac{dQ}{dt}
\end{equation}
where $P$ is the period of a radial pulsation mode in unit of days, $Q$ is the pulsation constant. For a specific mode, the Q value is an essential constant for all $\delta$ Scuti stars, hence the term $(1/Q)/(dQ/dt)$ is negligible as it is a very small quantity \citep{Breger2000}. The above relation is then reduced to as follows,

\begin{equation}
\frac{1}{P}\frac{dP}{dt}=-0.69\frac{dM_{bol}}{dt}-\frac{3}{T_{eff}}\frac{dT_{eff}}{dt}
\end{equation}

Stellar evolution leads to an increase in the period of most stars, from main-sequence stars to long-period evolved variable stars, with periods growing from 10$^{-10}$ yr$^{-1}$ to 10$^{-7}$ yr$^{-1}$ \citep{{Breger98}}. This period variation is observable and has been observed in some radial $\delta$ Scuti stars. \citet{{Breger98}} calculated the theoretical period variation of the radial fundamental modes of the 1.8$M_{\odot}$ model during the main-order and post-main-order evolution, and the results obtained are in agreement with the observed values. \citet{{2018ApJ...861...96X}} studied a HADS star VX Hya, by analyzing the period variation obtained from the $O-C$ analysis and the predicted values obtained from stellar evolution models, they found evolutionary effects can successfully explain the periodic change of this star.

We studied the period change of \target using the 17 quarters of $Kepler$ data. Due to the relatively large sampling interval of the $Kepler$ LC photometric observations, only 3 points available for each minimum times of the fit, the residuals of the results obtained using such data are relatively large. Therefore, we obtained the period change of the star using the same method as \citet{2017ampm.book.....B}. As can be seen, the $O-C$ plot is more diffuse than \citet{2017ampm.book.....B}. There are some possible factors contributing to the dispersion such as the $Kepler$ instrumental modulation, the number of radial modes detected
(e.g. the $O-C$ works worse for quadruple-mode than a double-mode), the presence of nonradial modes and, finally, some unresolved interactions between modes. In this sense it is worth noting that the frequency analysis performed with SigSpec was stopped after more than 200 iterations of the prewhitening cascade though the cumulative significance \citep[i.e. the joint probability distribution, for more details see][]{Reegen} was still higher than 8. While some of the extracted frequencies might be originated by aliases of high order harmonics and combinations, the presence of low amplitude nonradial modes appears to be clear. The presence of multiple components hampers the $O-C$ in time domain since it is not easy to estimate the time intervals between maxima, but in frequency domain we can resolve each component and measure their phases. The exploration of phase changes is, in this sense, a better way to study period changes than the classical $O-C$ in time domain. We still could expect some contribution to phase changes produced by interference when there is a high density of modes but this should not be significant otherwise.

\begin{figure}
\begin{center}
  \includegraphics[width=0.48\textwidth]{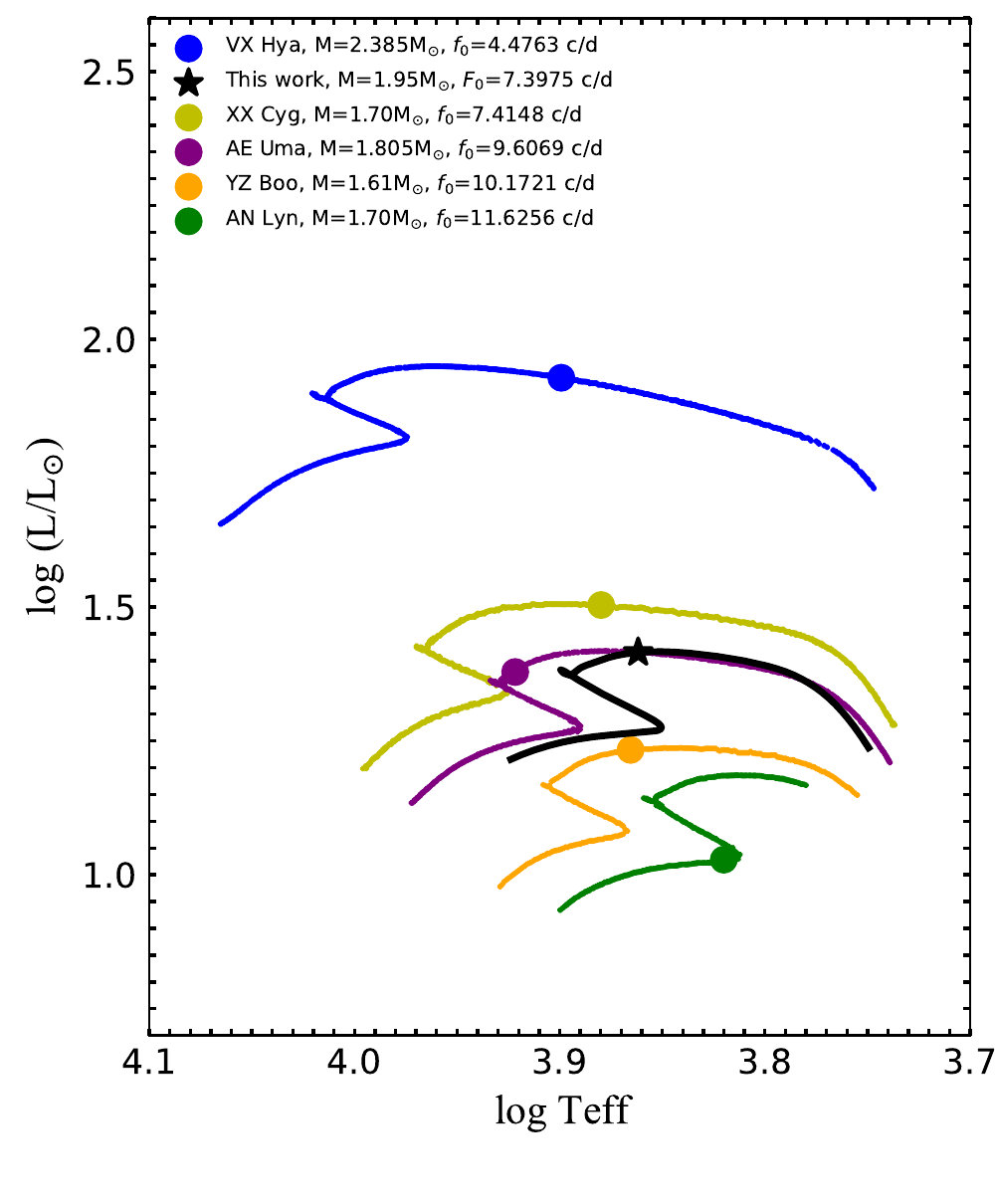}
  \caption{Evolution tracks for the best-fitting models of five well studied HADS and \target. See Table \ref{Tab5} for details.}
    \label{fig:bestmodel}
\end{center}
\end{figure}

\begin{deluxetable}{cccccccc}
\renewcommand\arraystretch{1.4}
\tabletypesize{\small}
\setlength\tabcolsep{1.5pt}
\tablewidth{0pc}
\tablenum{5}
\tablecaption{Observation-determined fundamental frequencies, period change rates and physical parameters from the best-fitting models of five HADS and \target. \label{Tab5}}
\tablehead{
\colhead{Star name}&
\colhead{$f_{0}$ (d$^{-1}$)}&
\colhead{$M/M_{\odot}$}&
\colhead{Age ($10^{9}\ \mathrm{years}$)}&
\colhead{[Fe/H]}&
\colhead{References}&
}
\startdata
VX\ Hya& 4.4763 &  2.385 & 0.43 & -0.2  & \citet{2018ApJ...861...96X} \\
\target& 7.3975 &  1.958  &1.016  & 0.02 &This paper\\
XX\ Cyg& 7.4148 &  1.70 & 0.9   & -0.49 & \citet{Yang2012} \\
YZ\ Boo& 9.6069 &  1.61  & 1.44  & -0.43 & \citet{Yang2018} \\
AN\ Lyn& 10.1721& 1.70  & 1.33  & 0.09  & \citet{Li2018} \\
AE\ UMa& 11.6256&  1.805 & 1.055 & -0.3  & \citet{Niu2017} \\
  \enddata
\end{deluxetable}

For HADS, the $O - C$ diagram is a powerful tool to investigate their period changes. According to \citet{Breger2000}, HADS can be divided into two types of increasing and decreasing periods. For instance, some HADS have an increasing period, i.e. YZ Boo \citep{Yang2018}; XX Cyg \citep{Yang2012}; GP And \citep{zhou2011}, etc., while others pulsate with a decreasing period, such as: BS Aqr \citep{2011Ap&SS.333..125B}; BE Lyn \citep{2011Ap&SS.333..125B}; DY Peg \citep{2003A&A...402..733D}, etc. Different values of period changes may suggest stars are in different stages of evolution. It is well established that $\delta$ Scuti stars (both Population I and II) observe much larger period variations than predicted by evolutionary models (e.g. \citealp{1995A&A...299..108R,Breger98,2001A&A...366..178R,2021MNRAS.504.4039B}). Thus \citet{2021MNRAS.504.4039B} concludes that the period changes observed in $\delta$ Scuti pulsators are not the result of stellar evolution, but may be related to the inherent non-linear excitation mechanism of high-amplitude pulsation modes, and the interactions of modes leading to modulated amplitudes and frequencies over time-scales of years and decades (e.g. \citealp{1985AcA....35....5D,1985AcA....35..229M,2014ApJ...783...89B,2016MNRAS.460.1970B}). For \target, the observed period changes is about two orders of magnitude larger than predicted by evolution theories. The possible reasons for this might be related to nonlinear mode interaction, but still need further investigation.

To compare with the previously discovered HADS studied with the stellar masses and evolutionary stages determined by asteroseismology, we selected five HADS (e.g. \citealp{Yang2012,Niu2017,Li2018,Yang2018,2018ApJ...861...96X}) and plot these positions along with the \target on the H$-$R diagram in Figure~\ref{fig:bestmodel}. Table \ref{Tab5} shows the observation-determined fundamental frequencies and the physical parameters from the best-fitting models for these five HADS and the average of the parameters obtained by our best models. Our results are consistent with \citet{2018ApJ...861...96X} which derives a trend that the lower the fundamental frequency, the more evolved the star is.

\begin{figure}
\begin{center}
  \includegraphics[width=0.5\textwidth]{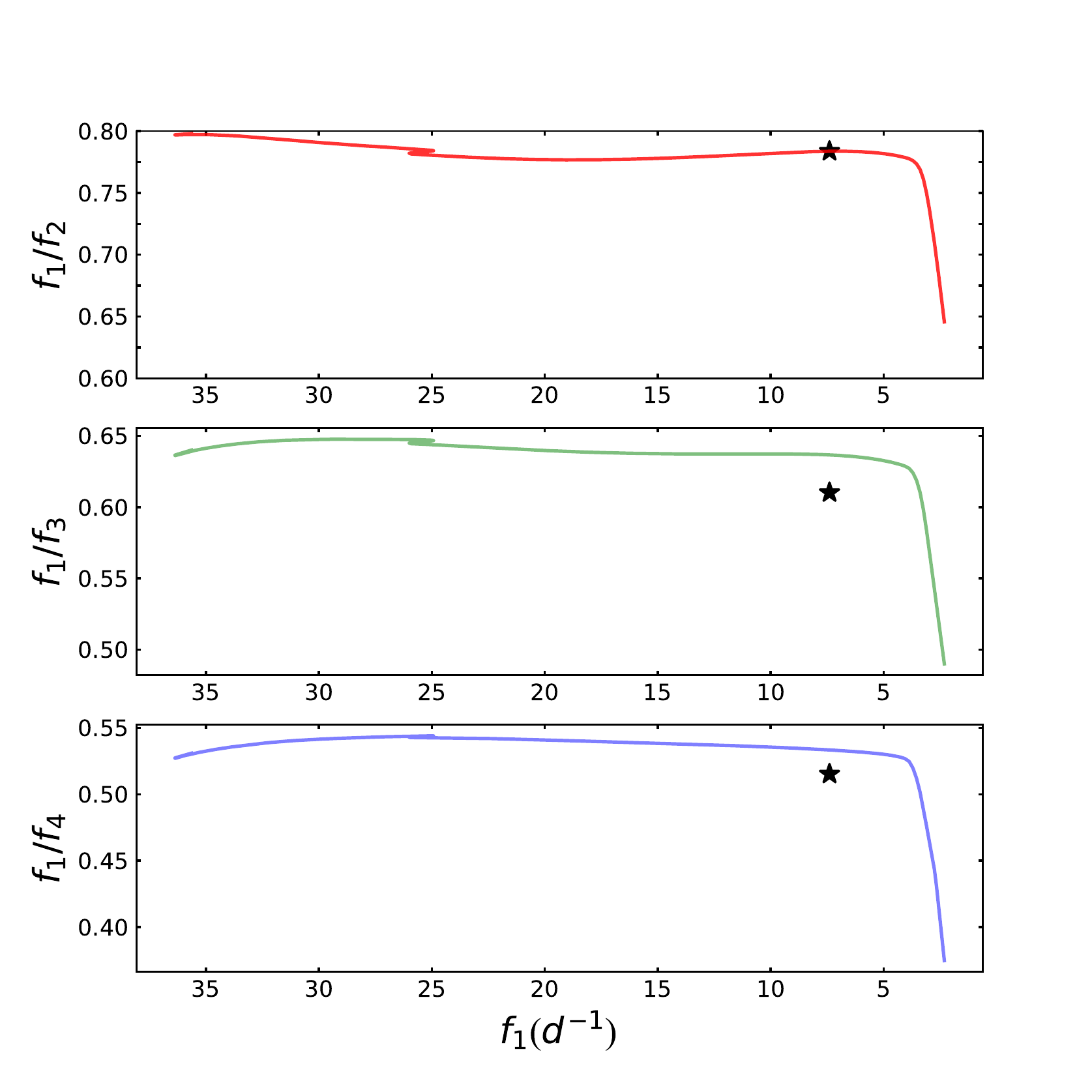}
  \caption{Petersen diagrams spanning from the ZAMS to the immediate post-main-sequence phase with $Z$ = 0.0007 and $M$ = 1.50$M/M_{\odot}$. The black star represent the observed ratio. From top to bottom, the ratio of the fundamental frequency to the first overtone, the ratio of the fundamental frequency to the second overtone, and the ratio of the fundamental frequency to the third overtone are shown successively.}
    \label{fig:poormetal}
\end{center}
\end{figure}

As can be seen from the Figure \ref{fig:modelsratio}, none of these models could match the observed frequency $f_{1}/f_{2}$ ratio of \target. Frequency uncertainties and the change of periods could account for no more than a difference of $10^{-4}$ between observed and theoretical values of frequency ratios. For the observed case where the ratio $f_{1}/f_{2}$ is greater than the ratio of the model, we have analysed several possible causes for this situation. The first possible reason is the effect of rotation on the ratio. \citet{2006A&A...447..649S} concludes that the period ratio $f_{1}/f_{2}$ increases with the increase of the rotation velocities by calculated the period ratios for different rotational velocities (Rotational Petersen Diagrams) and metallicities, and then compared with classic non-rotating ones. Even for slow rotators, the effect of rotation on the period ratio also can be significant \citep{2007A&A...474..961S}. Since \target is in the transition region from small amplitude to high amplitude, it should have a moderate rotation velocity, thus, the observed ratio is larger than the model, which may be caused by the rotation of the star. Lower metallicity has the effect of shifting period ratios towards slightly higher values for the same mass \citep{2005A&A...440.1097P}. So the second possible reason is that \target is a very poor metal star, when Petersen diagrams for extremely metal-poor star with $Z$ = 0.0007, the ratio of model $f_{1}/f_{2}$ could match the observed frequency ratio as shown in Figure \ref{fig:poormetal}. So the star is not excluded as a candidate for SX Phe. In this case, although the ratio of $f_{1}/f_{2}$ model is very close to the observed one, there is a large difference between the latter two ratios.

The mass-metallicity degeneracy of the fittings is notable, and similar to that found by \citet{2021MNRAS.504.4039B}. Due to the large uncertainties of $Z$ and the mass-metallicity degeneracy, we suggest that high-resolution spectroscopic observations of \target would not only help to accurately determine the metallicities that may break the degeneracies, but also provide other parameters such as effective temperature and rotation rate, thus further narrowing the parameter space of this star.

\section{Conclusions}

We have analyzed the pulsating behavior of \target using high-precision photometric observations from $Kepler$ mission, and 54 significant frequencies are detected, while four of them are independent frequencies, i.e. F0 = 7.3975 d$^{-1}$, F1 = 9.4397 d$^{-1}$, F2 = 12.1225 d$^{-1}$ and F3 = 14.3577 d$^{-1}$. The ratio of $f_{1}$ / $f_{2}$, $f_{1}$ / $f_{3}$ and $f_{1}$ / $f_{4}$ of \target are measured to be 0.783, 0.610 and 0.515, respectively, suggesting that this target could be a large-amplitude quadruple-mode $\delta$ Scuti star of the HADS group and average low-amplitude pulsators.

A different approach has been used to determine the $O-C$ through the phase modulation, the change of period $(1/P)~dP/dt$ is obtained resulting in $-1.14 \times 10^{-6}~\text{yr}^{-1}$ and $-4.48 \times 10^{-6}~\text{yr}^{-1}$ for F0 and F1 respectively. The dP/dt of the first overtone is consistent with that of the fundamental and is about two orders of magnitude larger than predicted by evolution theories. The possible reason might be related to nonlinear mode interaction but still need further investigation.

The stellar evolutionary models were constructed with different mass $M$ and metallicity $Z$ using MESA. Due to the effect of rotation on \target, we could not get a good match between the models and the observations. So we suggest high-resolution spectra is highly desired in the future, which would provide other parameters, and further narrow down the parameter space of this star. Thanks to the constraints provided by the four radial modes of oscillation of \target, we have shown in this work that $Kepler$ data can be provide a real-time picture of stellar evolution, thus opening a window to the development of ultra-precise stellar models.

\section*{Acknowledgments}

We thank the anonymous referee for the suggestive comments, which improved the manuscript. We would like to thank the $Kepler$ science team for providing such excellent data. This research is supported by the National Natural Science Foundation of China (grant No. U2031209 and 12003020). JPG acknowledge funding support from Spanish public funds for research from project PID2019-107061GB-C63 from the 'Programas Estatales de Generaci\'on de Conocimiento y Fortalecimiento Cient\'ifico y Tecnol\'ogico del Sistema de I+D+i y de I+D+i Orientada a los Retos de la Sociedad', and from the State Agency for Research through the "Center of Excellence Severo Ochoa" award to the Instituto de Astrof\'isica de Andaluc\'ia (SEV-2017-0709), all from the Spanish Ministry of Science, Innovation and Universities (MCIU).

Inlists used for our MESA analysis are available on Zenodo, at this\,\dataset[link]{https://zenodo.org/record/6442194\#.YlO0LjnYtjE}.

\appendix

\center{The additional independent non-radial pulsation mode frequencies identified in the LC $Kepler$ data of \target are provided in Tables~\ref{TabA}.}

\begin{deluxetable}{cccccc}[htbp]
\tabletypesize{\footnotesize}
\setlength\tabcolsep{8pt}
\linespread{1.0}
\tablewidth{0pc}
\tablenum{A}
\tablecaption{Additional independent frequencies extracted from the 4-yr LC $Kepler$ data of \target (i.e. F0, F1, F2, F3 and all their significant harmonics and combinations have been removed). \label{TabA}}
\tablehead{
\colhead{$f_{Si}$}   &
\colhead{Frequency (d$^{-1}$)}  &
\colhead{Amplitude (mmag)}      &
\colhead{S/N}            &
}
\startdata
1&	13.632342(1)&	2.53(4)&	186.1			\\
2&	9.79869(2)&	2.01(4)&	138.3		\\
3&	9.76036(2)&	0.95(4)&	65.5		\\
4&	17.19623(3)&	0.80(4)&	60.1		\\
5&	6.27140(3)&	0.57(4)&	34.7		\\
6&	2.40118(3)&	0.55(4)&	30.2		\\
7&	6.23481(4)&	0.55(4)&	33.8		\\
8&	15.62061(4)&	0.45(4)&	32.3 		\\
9&	3.76833(6)&	0.33(4)&	17.5		\\
10&	3.65501(6)&	0.30(4)&	16.2		\\
11&	0.35320(7)&	0.28(4)&	10.4  			\\
12&	0.71309(7)&	0.26(4)&	9.9  		\\
13&	0.36543(7)&	0.24(4)&	8.9  		\\
14&	15.43425(8)&	0.23(4)&	16.7  			\\
15&	0.35988(9)&	0.22(4)&	8.2  		\\
16&	4.9964(2)   &0.14(4)&	8.8     	\\
17&	12.4078(1)  &	 0.13(4)&	9.4     	  	\\
18&	12.5265(1)&	0.18(4)&	13.1 			\\
19&	4.1282(1)&	0.16(4)&	9.2  		\\
20&	3.0395(1)&	0.16(4)&	8.3  		\\
21&	21.029(1)&	0.17(4)&	12.8 		\\
22&	7.9564(1)&	0.14(4)&	8.2  		\\
23&	8.8715(1)   &	 0.13(4)&	8.7     		\\
24&	1.1261(1)&	0.13(4)&	6.9 		\\
25&	11.1093(1)&	0.12(4)&	8.7  		\\
26&	6.4504(1)&	0.12(4)&	7.0  		\\
27&	12.4475(1)&	0.11(4)&	8.2  		\\
28&	1.5463(2)&	0.10(4)&	5.8  		\\
29&	23.8921(2)&	0.10(4)&	7.3  		\\
30&	15.6200(2)&	0.10(4)&	7.5  		\\
31&	21.0669(2)&	0.10(4)&	8.2 		\\
   \enddata
\end{deluxetable}

\end{document}